\documentclass[draft,numberedheadings]{aipproc}
\layoutstyle{6x9}
\usepackage{natbib}

\def \hd {H$\delta$}

\def \caii {Ca~{\sc ii}}
\def \logm {$\log({\rm M/M}_\odot)$}

\begin{document}

\title{Quenching of Star Formation}

\classification{98.52.Cf, 98.62.Ai,02.50.Sk}
\keywords      {Galaxies, Spectroscopic surveys, star formation
history, post-starburst galaxies, robust statistics, semi-analytic
modelling} 

\author{Vivienne Wild}{
  address={Max-Planck Institut f\"{u}r Astrophysik, Karl-Schwarzschild Str. 1,
85741 Garching, Germany}
}

\author{Tamas Budav\'ari}{
  address={The Johns Hopkins
University, 3701 San Martin Drive, Baltimore, MD 21218, USA}
}

\author{J\'{e}r\'{e}my Blaizot}{
  address={Observatoire de Lyon, 9 avenue Charles Andr\'e, Saint-Genis Laval,
    F-69230, France}
}

\author{C.~Jakob Walcher}{
  address={Institut d'Astrophysique de Paris, UMR 7095, 98 bis Bvd Arago, 
75014 Paris, France}
}

\author{Peter H. Johansson}{
  address={Universit\"{a}ts-Sternwarte M\"{u}nchen, Scheinerstr. 1, D-81679
M\"{u}nchen, Germany}
}

\author{Gerard Lemson}{
  address={Astronomisches Rechen-Institut, Moenchhofstr. 12-14, 69120 Heidelberg, Germany}
}

\author{Gabriella de Lucia}{
  address={Max-Planck Institut f\"{u}r Astrophysik, Karl-Schwarzschild Str. 1,
85741 Garching, Germany}
}

\author{St\'{e}phane Charlot}{
  address={Institut d'Astrophysique de Paris, UMR 7095, 98 bis Bvd Arago, 
75014 Paris, France}
}

\begin{abstract}
In the last decade we have seen an enormous increase in the size and
quality of spectroscopic galaxy surveys, both at low and high
redshift. New statistical techniques to analyse large portions of
galaxy spectra are now finding favour over traditional index based
methods. Here we will review a new robust and iterative Principal
Component Analysis (PCA) algorithm, which solves several common issues
with classic PCA. Application to the 4000\AA\ break region of galaxies
in the VIMOS VLT Deep Survey (VVDS) and Sloan Digital Sky Survey
(SDSS) gives new high signal-to-noise ratio spectral indices easily
interpretable in terms of recent star formation history. In
particular, we identify a sample of post-starburst galaxies at
$z\sim0.7$ and $z\sim0.07$. We quantify for the first time the
importance of post-starburst galaxies, consistent with being
descendents of gas-rich major mergers, for building the red
sequence. Finally, we present a comparison with new low and high
redshift ``mock spectroscopic surveys'' derived from a Millennium Run
semi-analytic model.
\end{abstract} 

\maketitle

%%%%%%%%%%%%%%%%%%%%%%%%%%%%%%%%%%%%%%%%%%%%%%%%%%%%%%%%%%%%%%%%%%
\section{Motivation}
%%%%%%%%%%%%%%%%%%%%%%%%%%%%%%%%%%%%%%%%%%%%%%%%%%%%%%%%%%%%%%%%%%

The rest-frame optical spectrum of a galaxy contains a wealth of
information about its past and present star formation rate, chemical
evolution, dust content and the presence of an active galactic
nucleus. The advent of the Sloan Digital Sky Survey (SDSS) led us into
a new era in low redshift spectroscopic galaxy surveys, with the
number of unique, well calibrated, high quality, galaxy spectra
covering the full optical wavelength extent approaching $10^6$. At
high redshift, progress in the last decade has been similarly
significant.  Both the Vimos VLT Deep Survey (VVDS) and DEEP2 surveys,
which have released a large number of galaxy spectra to the public,
allow the same measurement of galaxy physical parameters at $z\sim1$
as routinely carried out at $z\sim0$.

At low redshift, the days of discreet classification of objects into
red/blue, elliptical/spiral etc. are over. The bimodality of the
galaxy population is well quantified in physical parameters such as
stellar mass and star formation rates. We are now in an era of
``galaxy population'' studies, in which each formerly distinct class
is thought of as part of a wider community. Rare classes of objects,
such as ``starburst'', ``green valley'' and ``post-starburst''
galaxies can be placed into a global picture of star formation
patterns. So-called ``transition'' galaxies are attracting great
interest, due to their importance for understanding the physical
mechanisms responsible for the global shut down in star formation, and
the build--up of the red sequence
\citep{2008MNRAS.387...79V,2008ApJ...675.1025S}. 

In this proceedings we will review recent work on applying principal
component analysis (PCA) to the 4000\AA\ break region of galaxy
spectra. This region provides constraints on the recent star formation
history of galaxies, important for identifying transition
galaxies. The 4000\AA\ break region is accessible to optical
spectroscopic surveys between $0<z<1$, making observations directly
comparable over half of the age of the Universe \cite{yanmei}, a time during which
the star formation habits of the galaxy population change
considerably.

%%%%%%%%%%%%%%%%%%%%%%%%%%%%%%%%%%%%%%%%%%%%%%%%%%%%%%%%%%%%%%%%%%
\section{Star formation histories from galaxy spectra}
%%%%%%%%%%%%%%%%%%%%%%%%%%%%%%%%%%%%%%%%%%%%%%%%%%%%%%%%%%%%%%%%%%

\begin{figure}
  \includegraphics[width=.4\textwidth]{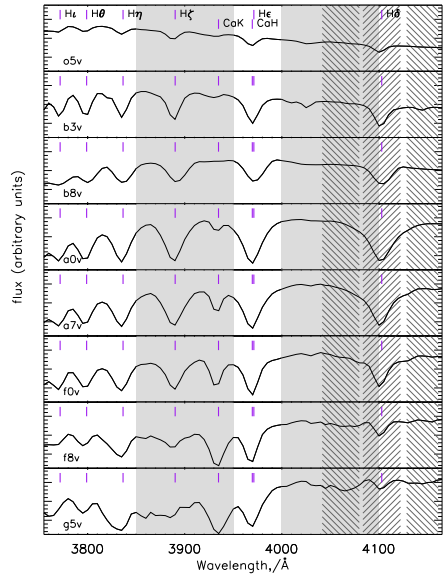} \caption{Example
  main sequence stellar spectra in the wavelength region of the
  4000\AA\ break. The stars are ordered by temperature, from hot to
  cold (i.e. shortest to longest main sequence lifetime). The 4000\AA\
  break strength (grey shaded areas) increases monotonically with
  decreasing temperature, while \hd\ equivalent width (hatched areas)
  decreases for the cooler stars, thus both provide powerful age
  indicators for young to intermediate age stellar populations.}
  \label{fig:stars}
\end{figure}

In the optical regime, the 4000\AA\ break region of the spectra
contains the greatest amount of information on the recent star
formation history of galaxies.  Figure \ref{fig:stars} illustrates
this point. With decreasing stellar temperature the 4000\AA\ break
strength increases, and the Balmer absorption lines first strengthen
and then weaken. A strong UV continuum is evident in the hottest
stars. Each stellar type has a characteristic main sequence lifetime,
$\sim$0.5\,Gyr and 3\,Gyr for A and F stars respectively. It is these
different lifetimes that allow us to measure the star formation
histories of stellar populations. 

PCA is an unsupervised multivariate statistical method traditionally
used to identify correlations in datasets. It has been applied
to astronomical spectral datasets for a variety of purposes
\citep[e.g.][]{1995AJ....110.1071C,1998ApJ...492...98G,2002MNRAS.333..133M,
2004AJ....128..585Y,2005MNRAS.358.1083W}.  PCA identifies correlated
features, such as the Balmer absorption lines, extracting them easily
as a single parameter. When applied to the 4000\AA\ break region of
galaxy spectra, PCA recovers the strength of the 4000\AA\ break as the
main axis of variation, which constrains the mean age of the stellar
population, or equivalently the specific star formation rate
\cite{2004MNRAS.351.1151B}. The second axis of variation is the Balmer
absorption line strength, which constrains the fraction of
intermediate age stars \cite{2003MNRAS.341...54K}. PCA also identifies
\caii(H\&K) as a clearly interpretable third axis.

In \citet{wild_psb} we developed new PCA-based spectroscopic indices
working in the 4000\AA\ break region, for the purpose of recovering
the recent star formation history of galaxies.We showed that, by
taking advantage of the entire Balmer series and continuum shape, a
dramatic improvement could be achieved over the traditionally used
H$\delta$ equivalent width. For this first application we chose to
create the PCA basis set using model galaxies from the
\citet{2003MNRAS.344.1000B} spectral synthesis models.  On the one hand, an
 oft-quoted benefit of PCA is that it can be applied directly to the
data, allowing the data to ``speak for themselves''. On the other
hand, direct application of classic PCA to modern galaxy data is
fraught with challenges caused by misclassified spectra, contamination
from night sky lines, regions of missing data and enormous
computational memory requirements. In the following section we will
discuss a new robust and recursive PCA algorithm developed in
\citet{2008arXiv0809.0881B}, designed specifically with large spectroscopic surveys
in mind.

%%%%%%%%%%%%%%%%%%%%%%%%%%%%%%%%%%%%%%%%%%%%%%%%%%%%%%%%%%%%%%%%%% 
\section{Robust and Iterative PCA}
%%%%%%%%%%%%%%%%%%%%%%%%%%%%%%%%%%%%%%%%%%%%%%%%%%%%%%%%%%%%%%%%%%

\begin{figure}
\includegraphics[width=.5\textwidth]{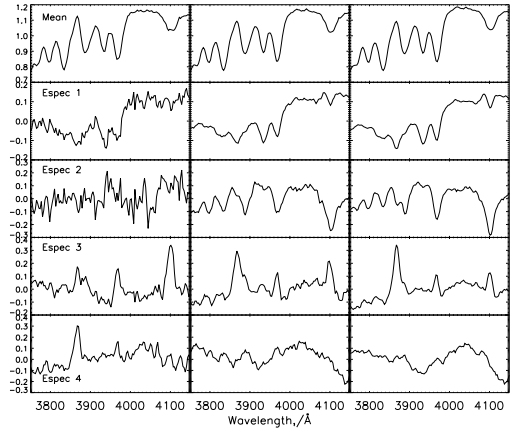}
\caption{The mean spectrum and top four eigenspectra for the VVDS
galaxies. {\it Left:} The result from classic PCA on 3485
spectra. {\it Center:} The result from classic PCA with iterative
removal of outliers. The final dataset contains 2675 spectra. {\it
Right:} The result from the new iterative-robust PCA algorithm.
}\label{fig:espec}
\end{figure}

In classical PCA, a set of eigenvectors are calculated through a matrix
decomposition of a single dataset. For very large samples, such as the
SDSS galaxy catalog, information is very much redundant in the statistical
sense, i.e. often the analysis of a smaller subset yields as good
results. Additionally, in most cases we seek only a small number of
eigenvectors associated with the largest eigenvalues, in classic PCA
the remaining vectors are computed in vain.

The first step of our new method is to formulate the problem within
the framework of a {\it data stream}, rather than the traditional {\it
data set}.  The eigensystem is recursively updated as new data are
input, with only the top $N$ eigenvectors calculated as
required. Convergence is controlled by a single parameter that sets
the effective sample size and can  easily be tuned to match the dataset
being analysed.

The second step is to make the algorithm robust to outliers. Classical
PCA simply minimises the square of the residual between the
eigensystem and input dataset, a statistical procedure inherently
susceptible to outliers. With small datasets it is possible to remove
the few ``obvious'' outliers by hand or within a few quick iterations
of classic PCA, a step which is both subjective and impractical for
modern large datasets. In the last few years, a number of improvements
have been proposed to overcome the issue of robustness in PCA, within
the framework of robust statistics \cite{maronna}. Rather than
minimising the square of the residuals, a robust function is
introduced to control the level of contamination tolerated from
outliers.

Figure \ref{fig:espec} presents a comparison between eigenspectra
created from classic PCA (left), those created using a traditional
workaround in which data are iteratively trimmed (center), and the new
iterative and robust method (right). The dataset is a sample of 3485
optical spectra from the VIMOS VLT Deep Survey
(VVDS)\cite{2005A&A...439..845L} which will be introduced in the
following section. A successful eigenbasis can be described as one
that does not introduce noise into the decomposition of individual
galaxy spectra, and in which the top few eigenspectra describe the
variance in the majority of the dataset. The classic PCA fails both of
these criteria: the eigenspectra are noisy, and the first two
component amplitudes show significant correlation for good quality
spectra. 

In this test case, the new eigenspectra are similar to those from the
trimmed PCA but small improvements are apparent. It is worth noting
that the PCA algorithm is completely independent of the order of the
bins: it has no spatial coherence.  Therefore, the fact that the new
eigenspectra are smoother is already an indication that they are more
robust.  The following section presents the first application
of indices derived using this method to a scientific problem.

%%%%%%%%%%%%%%%%%%%%%%%%%%%%%%%%%%%%%%%%%%%%%%%%%%%%%%%%%%%%%%%%%%
\section{Post-starburst galaxies at $z\sim0.7$}
%%%%%%%%%%%%%%%%%%%%%%%%%%%%%%%%%%%%%%%%%%%%%%%%%%%%%%%%%%%%%%%%%%

\begin{figure}
\begin{minipage}{\textwidth}
\begin{minipage}{\textwidth}
\includegraphics[width=.45\textwidth]{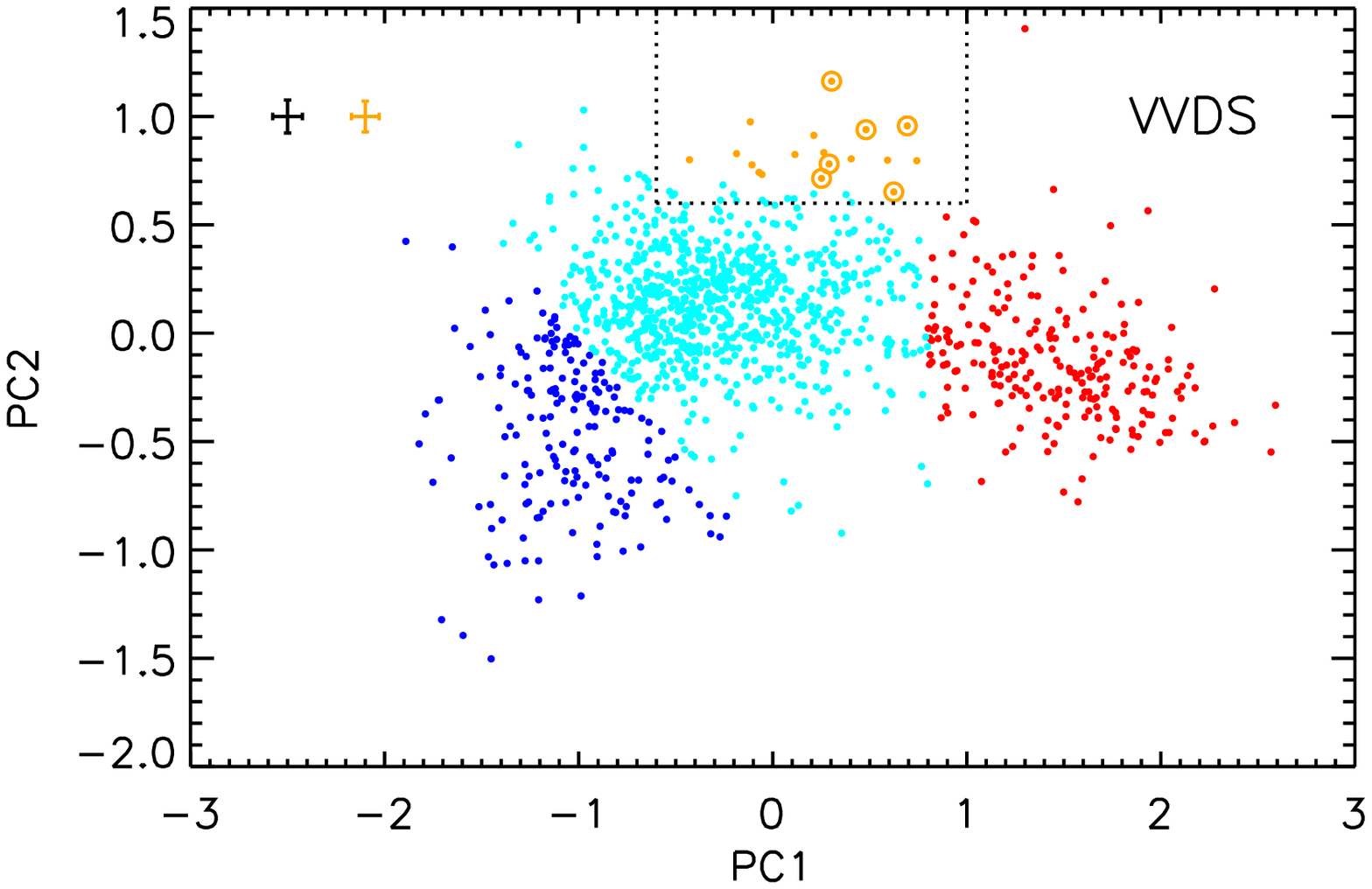}
\includegraphics[width=.45\textwidth]{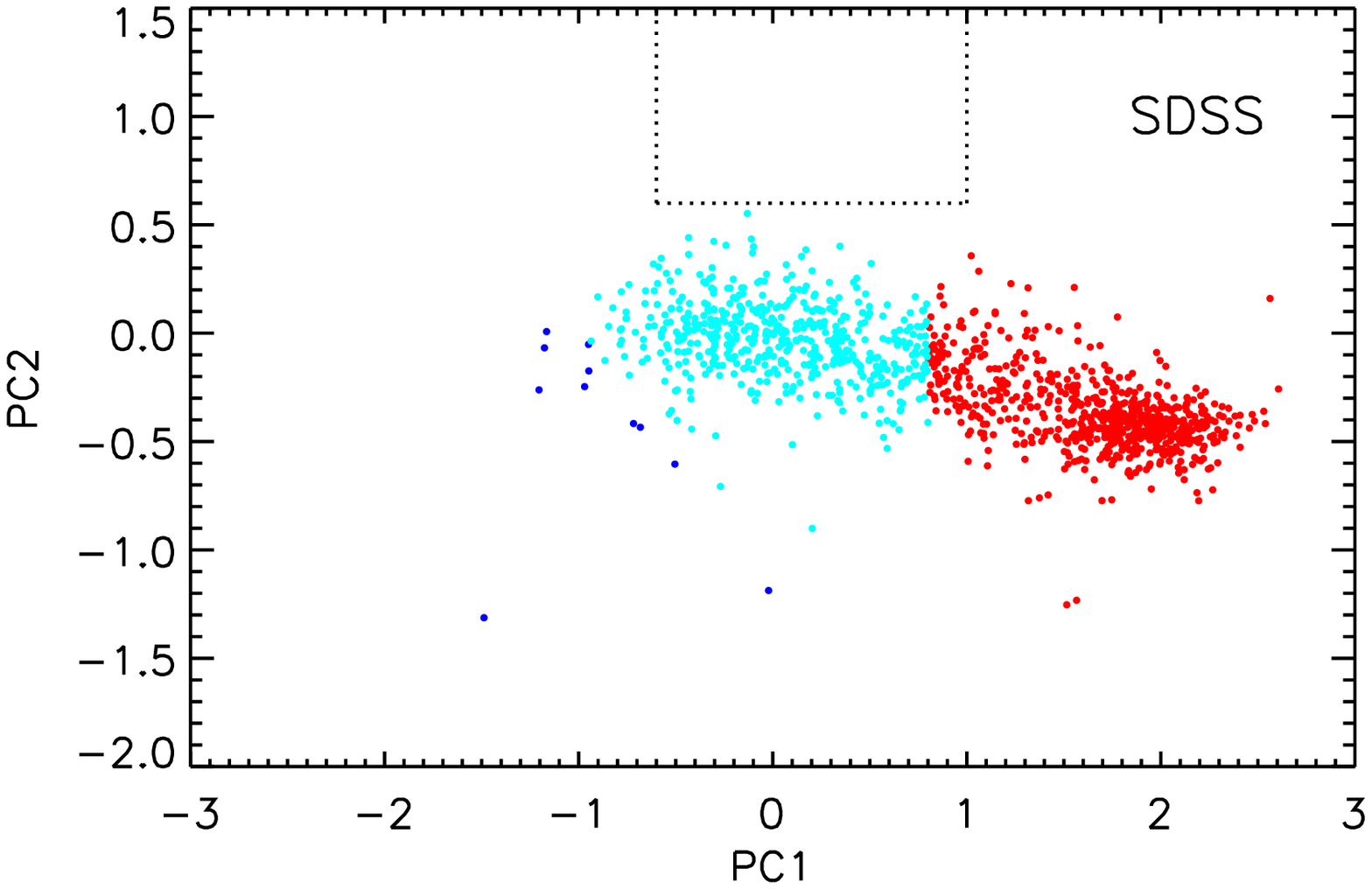}\\
\end{minipage}
\begin{minipage}{\textwidth}
\includegraphics[width=.45\textwidth]{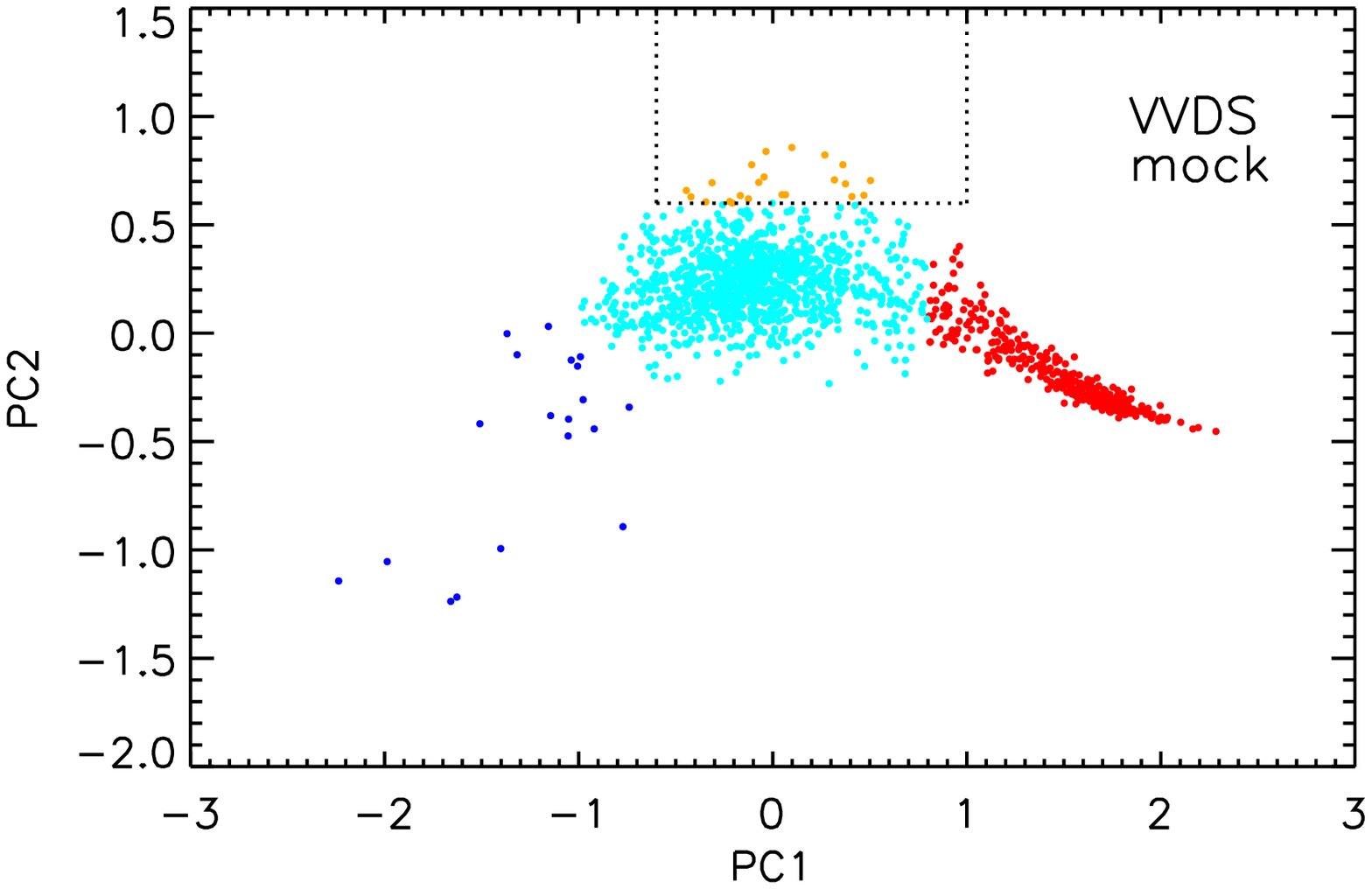}
\includegraphics[width=.45\textwidth]{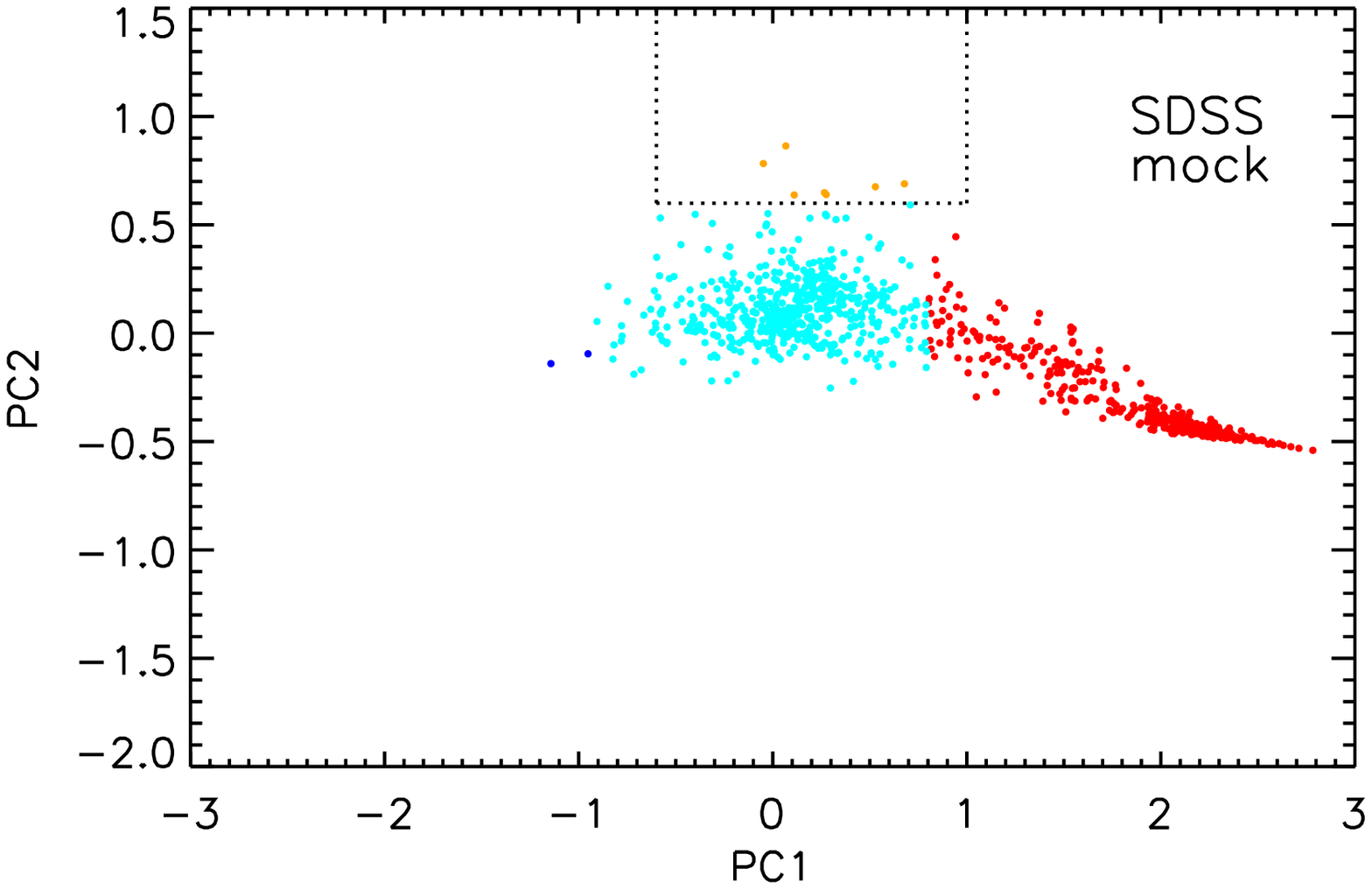}
\end{minipage}
\end{minipage}

\caption{The first two principal components (PCs) for
different samples of $\sim$1250 galaxies. The samples have been split
  into quiescent (red), starforming (cyan), star-bursting (blue) and
  post-starburst (orange) classes. Post-starburst galaxies are defined
  to lie within the dotted box indicated by at least $1\sigma$. {\it
  Top Left:} VVDS galaxies with $0.5<z<1.0$. Post-starburst galaxies
  with SSFR$<10^{-11}$/yr are circled. The median errors of the whole
  sample (black) and the post-starburst galaxies alone (orange) are
  shown in the top left. {\it Top right:} a comparison SDSS low
  redshift sample with $0.05<z<0.1$ and \logm$>9.75$. The same number
  of galaxies as in the VVDS sample have been randomly selected for
  illustration purposes. The lower panels depict {\it work in
  progress} to recreate both VVDS and SDSS spectroscopic surveys from
  the Millennium Run  semi-analytic model of
  \citet{2007MNRAS.375....2D}.}\label{fig:pca}
\end{figure}

The VIMOS VLT Deep survey (VVDS) is a deep spectroscopic redshift
survey, targetting objects with apparent magnitudes in the range of
$17.5\le I_{AB}\le24$ \cite{2005A&A...439..845L}. The survey is unique
for high redshift galaxy surveys in having applied no colour cuts,
yielding a particularly simple selection function, making it a very
attractive dataset for statistical studies of the high redshift galaxy
population. In this work we make use of the spectra from the publicly
available first epoch data release of the VVDS-0226-04 (VVDS-02h)
field. The spectra have a relatively low resolution of $R\!=\!227$ and
observed frame wavelength range of $\sim$5500-9500\AA. The first epoch
public data release contains 8981 spectroscopically observed objects
in the VVDS-02h field, from which we select 1246 with secure redshifts
in the range $0.5<z<1.0$ and a per-pixel signal-to-noise ratio (SNR) greater
than 6.5.

The new robust and iterative PCA method was applied to these spectra
to create the eigenspectra directly from the VVDS spectra. In the top left
panel of Figure \ref{fig:pca} we show the distribution of the first
two principal component amplitudes, PC1 and PC2, for this sample of
VVDS galaxies. The primary division of our sample is into
``quiescent'' and ``star-forming'' galaxies on the right and left. To
the bottom left, ``starburst'' galaxies are found with very blue
continua and very strong emission lines. To the top centre we find the
``post-starburst'' galaxies, with stronger Balmer absorption than
expected for their 4000\AA\ break strength. In the top right panel we show
a comparison sample derived from the SDSS DR6 catalog, 1246 galaxies
with $0.05<z<0.1$ have been selected randomly for illustration
purposes\footnote{We note that these samples are {\it not} mass
limited, and therefore the completeness increases for a given stellar
mass from right to left across the diagrams. However, the completeness
limits of the two samples are similar, and the left and right panels
may be directly compared.}. There are some clear differences between
the high and low redshift samples. Firstly, the entire galaxy
population noticeably ages (increases in PC1) as redshift decreases,
the red sequence builds and the strong starburst galaxies are no
longer visible in the SDSS survey. Secondly, the scatter in both the
blue and red sequence drops substantially with decreasing redshift,
even when the higher SNR of the SDSS spectra is accounted for. The
SDSS shows a clear blue sequence, whereas the VVDS shows a ``blue
cloud''. 

In \citet{vvds_psb} we compare with SPH simulations of galaxy mergers
\cite{2008arXiv0802.0210J} to show that the strong post-starburst
(PSB) galaxies found in the VVDS and SDSS surveys are consistent with
being the descendents of gas rich major mergers. The inclusion of
black hole feedback does not greatly alter the evolution of the
simulated merger remnants through the post-starburst phase. Starburst
mass fractions must be larger than $\sim5-10$\% and decay times
shorter than $\sim10^8$ years for post-starburst spectral signatures to be
observed in the simulations.

In the VVDS survey, we find 16 PSB galaxies above our mass
completeness limit of \logm$>9.75$. These correspond to a number
density of $1\times 10^{-4}$Mpc$^{-3}$. Summing the mass of the 5 PSB
galaxies which have completely ceased the formation of stars, and are
therefore most likely to enter the red sequence, we measure a mass
flux from blue to red sequence through the post-starburst phase of
$\dot{\rho}_{A
\rightarrow Q, PSB} = 0.0038^{+0.0004}_{-0.001}$
M$_\odot$/Mpc$^3$/yr. Comparison with \citet{2007A&A...476..137A}
shows that this accounts for $38^{+4}_{-11}$\% of the growth rate of
the red sequence at $z\sim0.7$.

We find a very strong redshift evolution of these strong
post-starburst galaxies: the number density is 200 times lower at
$z\sim0.07$ than at $z\sim0.7$. In the redshift range $0.05<z<0.1$,
only 3 equivalently strong post-starburst galaxies are found in the
SDSS-DR6 catalogue above our mass completeness limit of
\logm$>9.75$. The strength of this evolution suggests that a combination of
effects are responsible: declining merger rates
\cite{2008arXiv0807.2578D}, declining gas fractions and increasing
disc dynamical timescales, leading to increasing burst durations and
weakening burst strengths.

\subsection{Comparison with cosmological simulations}

In the lower panels of Figure \ref{fig:pca} we present recent results
from work-in-progress to extract mock spectroscopic catalogues from
the Millennium Run \cite{2005Natur.435..629S} semi-analytic model
(SAM) of
\citet{2007MNRAS.375....2D}.  Mock magnitude-limited catalogues are
created using the MoMaf software \cite{2005MNRAS.360..159B}, and the
star formation and metallicity history of the SAM galaxies are
combined with the \citet{2003MNRAS.344.1000B} spectral synthesis
models to create integrated spectral energy distributions for each SAM
galaxy. Prescriptions for nebular emission lines and two component
dust attenuation are included \citep{2000ApJ...539..718C}. The
resulting spectra can then be analysed in an entirely equivalent
manner to the real data. In the case of Figure \ref{fig:pca}, the SAM
galaxy spectra have been projected onto the robust eigenspectra
created from the VVDS galaxy spectra as described above.

Comparing the upper (real) and lower (mock) panels of Figure
\ref{fig:pca} we can be encouraged by the agreement between
the SAM and real Universes. In particular, the overall ageing of the
stellar populations is well reproduced, together with the decreasing
scatter in the blue sequence as the Universe ages. Some shortcomings
of the models are evident, and are currently being fully
investigated. The figures make a strong case that the application of
multivariate statistics to data compression and visualisation of large
and complex datasets will lead to new insights into the physical
processes responsible for driving the changes in star formation in
galaxies over time.

%%%%%%%%%%%%%%%%%%%%%%%%%%%%%%%%%%%%%%%%%%%%%%%%%%%%%%%%%%%%%%%%%%
\section{Conclusions}
%%%%%%%%%%%%%%%%%%%%%%%%%%%%%%%%%%%%%%%%%%%%%%%%%%%%%%%%%%%%%%%%%%
In this proceedings we have gathered together recent and ongoing work
by the authors, to quantify the recent star formation history of
galaxies through use of the 4000\AA\ break region of galaxy
spectra. Through a new robust PCA algorithm, we have derived high SNR
spectroscopic indices from the high redshift VVDS galaxy survey. These
indices combine the Balmer absorption line series and shape of the
spectra to reveal the relative fractions of stellar types. We have
shown how these indices can be applied to current scientific
questions, by the identification of post-starburst galaxies. Finally,
we have presented new work on extracting full spectral energy
distributions for Millennium Run SAM galaxies, including prescriptions
for dust attenuation and nebular emission. We have shown how the data
compression and visualisation properties of PCA help us to compare
directly these modern large cosmological simulations with the modern,
and equally large, spectroscopic galaxy surveys.

%%%%%%%%%%%%%%%%%%%%%%%%%%%%%%%%%%%%%%%%%%%%%%%%
%% BACKMATTER
%%%%%%%%%%%%%%%%%%%%%%%%%%%%%%%%%%%%%%%%%%%%%%%%

%\begin{theacknowledgments}
%MAGPop and PI Guinevere Kauffmann
%\end{theacknowledgments}



\bibliographystyle{aipproc}   % if natbib is available

\bibliography{refs_all}

\begin{thebibliography}{22}
\expandafter\ifx\csname natexlab\endcsname\relax\def\natexlab#1{#1}\fi
\providecommand{\enquote}[1]{``#1''}
\expandafter\ifx\csname url\endcsname\relax
  \def\url#1{\texttt{#1}}\fi
\expandafter\ifx\csname urlprefix\endcsname\relax\def\urlprefix{URL }\fi
\providecommand{\eprint}[2][]{\url{#2}}

\bibitem[{van den Bosch} et~al.(2008)]{2008MNRAS.387...79V}
F.~C. {van den Bosch}, D.~{Aquino}, X.~{Yang}, {et~al.}, \emph{\mnras} \textbf{387},
  79--91 (2008)

\bibitem[{Silverman} et~al.(2008)]{2008ApJ...675.1025S}
J.~D. {Silverman}, V.~{Mainieri}, B.~D. {Lehmer}, {et~al.}, \emph{\apj}
  \textbf{675}, 1025--1040 (2008)

\bibitem[{Chen} et~al.(2008)]{yanmei}
Y.-M. {Chen}, V.~{Wild}, G.~A.~M. {Kauffmann}, {et~al.}, \emph{ArXiv
  e-prints} \textbf{0808:3683}  (2008).

\bibitem[{Connolly} et~al.(1995)]{1995AJ....110.1071C}
A.~J. {Connolly}, A.~S. {Szalay}, M.~A. {Bershady}, A.~L. {Kinney}, and
  D.~{Calzetti}, \emph{\aj} \textbf{110}, 1071, (1995)

\bibitem[{Glazebrook} et~al.(1998)]{1998ApJ...492...98G}
K.~{Glazebrook}, A.~R. {Offer}, and K.~{Deeley}, \emph{\apj} \textbf{492},
  98 (1998).

\bibitem[{Madgwick} et~al.(2002)]{2002MNRAS.333..133M}
D.~S. {Madgwick}, O.~{Lahav}, I.~K. {Baldry}, {et al. (The 2dFGRS Team),},
  \emph{\mnras} \textbf{333}, 133--144 (2002).

\bibitem[{Yip} et~al.(2004)]{2004AJ....128..585Y}
C.~W. {Yip}, A.~J. {Connolly}, A.~S. {Szalay}, {et~al.}, \emph{\aj}
  \textbf{128}, 585--609 (2004)

\bibitem[{Wild} and {Hewett}(2005)]{2005MNRAS.358.1083W}
V.~{Wild}, and P.~C. {Hewett}, \emph{\mnras} \textbf{358}, 1083--1099 (2005).

\bibitem[{Brinchmann} et~al.(2004)]{2004MNRAS.351.1151B}
J.~{Brinchmann}, S.~{Charlot}, S.~D.~M. {White}, {et~al.}, \emph{\mnras} \textbf{351}, 1151--1179
  (2004).

\bibitem[{Kauffmann} et~al.(2003)]{2003MNRAS.341...54K}
G.~{Kauffmann}, T.~M. {Heckman}, S.~D.~M. {White}, {et~al.,}, \emph{\mnras}
  \textbf{341}, 54--69 (2003).

\bibitem[{Wild} et~al.(2007)]{wild_psb}
V.~{Wild}, G.~{Kauffmann}, T.~{Heckman}, {et~al.}, \emph{\mnras}
  \textbf{381}, 543--572 (2007)

\bibitem[{Bruzual} and {Charlot}(2003)]{2003MNRAS.344.1000B}
G.~{Bruzual}, and S.~{Charlot}, \emph{\mnras} \textbf{344}, 1000--1028 (2003).

\bibitem[{Budavari} et~al.(2008)]{2008arXiv0809.0881B}
T.~{Budavari}, V.~{Wild}, A.~S. {Szalay}, L.~{Dobos}, and C.-W. {Yip},
   \emph{ArXiv
  e-prints} \textbf{0809:0881} (2008)

\bibitem[{Maronna} et~al.(2006)]{maronna}
R.~{Maronna}, R.~{Martin}, and V.~{Yohai}, \emph{{Wiley Series in Probability
  and Statistics}}  (2006).

\bibitem[{Le F{\`e}vre} et~al.(2005)]{2005A&A...439..845L}
O.~{Le F{\`e}vre}, G.~{Vettolani}, B.~{Garilli}, and {et~al.}, \emph{\aap}
  \textbf{439}, 845--862 (2005)

\bibitem[{De Lucia} and {Blaizot}(2007)]{2007MNRAS.375....2D}
G.~{De Lucia}, and J.~{Blaizot}, \emph{\mnras} \textbf{375}, 2--14 (2007)

\bibitem[{Wild} et~al.(2008)]{vvds_psb}
V.~{Wild}, C.~J. {Walcher}, P.~{Johansson}, {et~al.}, \emph{MNRAS submitted}  (2008).

\bibitem[{Johansson} et~al.(2008)]{2008arXiv0802.0210J}
P.~H. {Johansson}, T.~{Naab}, and A.~{Burkert}, \emph{ArXiv e-prints}
  \textbf{0802.0201} (2008)

\bibitem[{Arnouts} et~al.(2007)]{2007A&A...476..137A}
S.~{Arnouts}, C.~J. {Walcher}, O.~{Le F{\`e}vre}, and {et~al.}, \emph{\aap}
  \textbf{476}, 137--150 (2007)

\bibitem[{de Ravel} et~al.(2008)]{2008arXiv0807.2578D}
L.~{de Ravel}, O.~{Le F{\`e}vre}, L.~{Tresse}, and {et~al.}, \emph{ArXiv
  e-prints} \textbf{0807:2578} (2008)

\bibitem[{Springel} et~al.(2005)]{2005Natur.435..629S}
V.~{Springel}, S.~D.~M. {White}, A.~{Jenkins}, and {et~al.}, \emph{\nat}
  \textbf{435}, 629--636 (2005)

\bibitem[{Blaizot} et~al.(2005)]{2005MNRAS.360..159B}
J.~{Blaizot}, Y.~{Wadadekar}, B.~{Guiderdoni}, {et~al.}, \emph{\mnras}
  \textbf{360}, 159--175 (2005)

\bibitem[{Charlot} and {Fall}(2000)]{2000ApJ...539..718C}
S.~{Charlot}, and S.~M. {Fall}, \emph{\apj} \textbf{539}, 718--731 (2000).

\end{thebibliography}

\end{document}